\documentstyle[preprint,tighten,aps,epsf]{revtex}

\newcommand{\qsla}{q\hspace{-0.19cm}/}

\newcommand{\esla}{\varepsilon\hspace{-0.2cm}/}

\begin{document}
\preprint{MKPH-T-97-13}
\draft
\title{Generalized polarizabilities of the nucleon studied in the linear
 sigma model (II)}
\author {A. Metz and D. Drechsel}
\address{Institut f\"ur Kernphysik, Johannes Gutenberg-Universit\"at Mainz, 
J. J. Becher-Weg 45, D-55099 Mainz, Germany}
\date{\today}
\maketitle

\begin{abstract}
In a previous paper virtual Compton scattering off the nucleon has been
investigated in the one--loop approximation of the linear sigma model 
in order to determine the 3 scalar generalized polarizabilities.
We have now extended this work and calculated the 7 vector polarizabilities
showing up in the spin--dependent amplitude of virtual Compton scattering.
The results fulfill 3 model--independent constraints recently derived.
Compared to the constituent quark model there exist enormous differences
for some of the vector polarizabilities.
At vanishing three--momentum of the virtual photon, the analytical results 
of the sigma model and of chiral perturbation theory can be related.
The influence of the $\pi^{0}$ exchange in the $t$ channel has been 
discussed in some detail.
Besides, the vector polarizabilities determine 2 linear combinations
of the third order spin--polarizabilities appearing in real Compton 
scattering.
\end{abstract}

\section{Introduction}
\label{kap_1}

Recently, it has been proposed to study the structure of the proton 
by the reaction $p(e,e'p)\gamma$.
The reason is that this process contains, in addition to electron 
bremsstrahlung (Bethe--Heitler scattering), the amplitude of virtual 
Compton scattering (VCS) off the proton, 
$\gamma^{\ast} + p \to \gamma + p$.
In general, Compton scattering as a two step process allows to extract
information on the excitation spectrum of the target. 

An overview of the various aspects of VCS can be found in Ref. 
\cite{Proc_clermont_96}.
VCS is of particular interest below pion production threshold, where
the information about the excited states of the nucleon can be parametrized 
by means of the generalized electromagnetic polarizabilities.
These generalized polarizabilities emerge as coefficients if the non Born 
amplitude of VCS is expanded in terms of the final photon energy 
$\omega' = 0$.
As has been demonstrated by Guichon et al. \cite{Guichon_95}, the 
leading term of such an expansion contains 10 generalized polarizabilities,
3 of them in the spin--independent amplitude (scalar polarizabilities) 
and 7 in the spin--flip amplitude (vector polarizabilities).
In contrast to real Compton scattering (RCS), the generalized polarizabilities
in VCS are functions depending on the four--momentum transfer $Q^{2}$. 

At present, three experimental programs are under way to determine
the generalized polarizabilities.
At MAMI, data have been taken at $Q^{2} = 0.33 \, \text{GeV}^{2}$ 
\cite{MAMI_proposal}.
There also exist plans to investigate VCS at even lower values of $Q^{2}$
at MIT--Bates \cite{BATES_proposal}, while the activities at Jefferson Lab
will concentrate on the region of higher $Q^{2}$ \cite{CEBAF_proposal}.
In an unpolarized experiment only 3 independent linear combinations
of the polarizabilities can be determined \cite{Guichon_95,Drechsel_97}.
A separate measurement of all polarizabilities is only possible by
use of double polarization observables, e.g., the reaction 
$p(\vec{e},e'\vec{p}\,)\gamma$ \cite{Vanderhaeghen_97b}.
Moreover, a careful treatment of radiative corrections is unavoidable 
as their contributions are comparable to the polarizability
effects \cite{Vanderhaeghen_97a}. 

Regarding the expansion in $\omega'$ for RCS, higher order terms can not
be neglected at photon energies larger than about $80 \, \text{MeV}$.
In order to guarantee the dominance of the leading order term, the energy 
of the real photon should also satisfy $\omega' \ll E - m_{N}$, 
where $E$ is the energy of the initial nucleon in the {\it cm} system 
and $m_{N}$ the nucleon mass \cite{Drechsel_96}, i.e., $\omega'$ has to
be much smaller than the three--momentum $|\, \vec{q} \,|$ of the 
virtual photon (see section \ref{kap_2}).
If $\omega'$ and $|\, \vec{q} \,|$ are of the same order of magnitude,
the amplitude has to be expanded in both variables 
(see Ref. \cite{Fearing_96} for the case of a spin 0 target).
The extension of this work to the nucleon leads to 36 independent 
low energy constants entering the non--Born VCS amplitude to fifth
order \cite{Drechsel_97}.  
While it is certainly hopeless to measure these parameters,
we want to emphasize the importance of a reliable estimate of the 
higher order terms, except for very small values of $\omega'$. 

Recently, it has been shown that the combined symmetry of charge 
conjugation and nucleon crossing results in unexpected relations between 
the VCS multipoles, beyond the usual constraints of parity and angular 
momentum conservation \cite{Drechsel_96,Drechsel_97}.
As a consequence, only 2 independent scalar and 4 independent vector 
polarizabilities exist. 

The generalized polarizabilities of the proton were first predicted
by use of the constituent quark model (CQM) \cite{Guichon_95,Liu_96}. 
Further calculations in an effective Lagrangian model \cite{Vanderhaeghen_96},
the linear sigma model (LSM) \cite{Metz_96a}, chiral perturbation theory 
(ChPT) \cite{Hemmert_96a,Hemmert_96b}, and the Skyrme model \cite{Kim_97}
focused on the scalar polarizabilities and the spin--independent VCS 
amplitude.
In the meantime, the vector polarizabilities have been evaluated on the 
basis of ChPT \cite{Hemmert_97,Hemmert_97a} and in a coupled--channel 
unitary model \cite{Korchin_97}. 

The present contribution completes our earlier work on the LSM 
\cite{Metz_96a} by presenting the results for the vector polarizabilities.
Since the model is chirally invariant, we expect similar results as in
ChPT for $Q^{2} \to 0$, and in addition a reasonable estimate for the 
$Q^{2}$ dependence of the polarizabilities.

\setcounter{equation}{0}
\section{Kinematics}
\label{kap_2}

For convenience we briefly repeat the definition of the kinematical 
variables for the VCS reaction \cite{Metz_96a,Metz_96b} 
 \begin{equation} \label{gl_1}
 \gamma^{\ast} (q^{\mu} , \varepsilon^{\mu}) + N (p^{\mu} , S^{\mu}) \to
 \gamma (q'^{\mu},\varepsilon'^{\mu}) + N' (p'^{\mu} , S'^{\mu}) \, ,
 \quad \textrm{with} \;\; q^{2} = -Q^{2} < 0 \, , \;\; q'^{2} = 0 \, .
 \end{equation}
The {\it cm} energies and three--momenta of the involved particles are denoted
by
$q^{\mu} = (\omega,\vec{q} \,) \,, \; 
q'^{\mu} = (\omega',\vec{q}\,') \,, \;
p^{\mu} = (E,-\vec{q} \,) \,, \; 
p'^{\mu} = (E',-\vec{q}\,')$.
The polarization vectors of both photons can be chosen to be spacelike, 
i.e., $\varepsilon^{\mu} = (0,\vec{\varepsilon} \,)$ and
$\varepsilon'^{\mu} = (0,\vec{\varepsilon}\,')$,
noting that the time--like component of the virtual photon can be
eliminated by current conservation.
While the outgoing photon is purely transverse (helicity $\lambda = \pm 1$), 
the incoming photon has also longitudinal polarization ($\lambda = 0$),
denoted by $\varepsilon^{\mu} (\lambda = 0) = (0,\hat{q})$.
In the following we also use the definitions
 \begin{eqnarray} \label{gl_2}
 \omega_{0} & = & \omega |_{\omega' = 0} = m_{N} - E
 = m_{N} - \sqrt{\bar{q}^{2} + m_{N}^{2}} \,,
 \\ \label{gl_3}
 Q_{0}^{2} & = & Q^{2}|_{\omega' = 0}
 = -2 m_{N} \omega_{0} \,,
 \end{eqnarray}
with $\bar{q} \equiv |\, \vec{q} \,|$.

With the convention of Bjorken and Drell \cite{Bjorken_64} for the 
invariant matrix element ${\cal{M}}$ we define the scattering
amplitude $T$ by 
\begin{equation} \label{gl_4}
T = - \frac{i}{4\pi} {\mathcal{M}}
= - \frac{i}{4\pi} \varepsilon_{\mu}'^{\ast} H^{\mu\nu} \varepsilon_{\nu} \,.
\end{equation}
As the Compton tensor $H^{\mu\nu}$ consists of 12 basis elements
\cite{Tarrach_75,Drechsel_97}, $T$ can be decomposed into 12 independent 
amplitudes.
Following Guichon et al. \cite{Guichon_95}, we parametrize the longitudinal 
and the transverse part of $T$ in Pauli space,
\begin{eqnarray} \label{gl_5}
T(\lambda = 0) & = & 
\chi_{f}^{\dagger} \big[ 
\vec{\varepsilon}\,'^* \cdot \hat{q} \, a^{l}
+ i \, \vec{\varepsilon}\,'^* \cdot ( \hat{q} \times \hat{q}' ) \, 
\vec{\sigma} \cdot \hat{e}_{x} \, b_{1}^{l}
\nonumber\\
& & \quad \; + i \, \vec{\varepsilon}\,'^* \cdot \hat{q} \,
\vec{\sigma} \cdot \hat{e}_{y} \, b_{2}^{l}
+ i \, \vec{\varepsilon}\,'^* \cdot ( \hat{q} \times \hat{q}' ) \,
\vec{\sigma} \cdot \hat{e}_{z} \, b_{3}^{l} \big] \chi_{i} \, ,
\\ 
& & \vphantom{a} 
\nonumber \\ \label{gl_6}
T(\lambda = \pm 1) & = & 
\chi_{f}^{\dagger} \big[ 
\vec{\varepsilon}\,'^* \cdot \vec{\varepsilon} \, a^{t}
+ \vec{\varepsilon}\,'^* \cdot \hat{q} \,
\vec{\varepsilon} \cdot \hat{q}' \, a^{t \prime}
\nonumber\\
& & \quad \; + i \, \vec{\varepsilon}\,'^* \cdot \hat{q} \,
\vec{\varepsilon} \cdot ( \hat{q} \times \hat{q}' ) \,
\vec{\sigma} \cdot \hat{e}_{x} \, b_{1}^{t}
+ i \, \vec{\varepsilon} \cdot \hat{q}' \,
\vec{\varepsilon}\,'^* \cdot ( \hat{q} \times \hat{q}' ) \,
\vec{\sigma} \cdot \hat{e}_{x} \, b_{1}^{t \prime}
\nonumber\\
& & \quad \; + i \, \vec{\varepsilon}\,'^* \cdot \vec{\varepsilon} \,
\vec{\sigma} \cdot \hat{e}_{y} \, b_{2}^{t}
+ i \, \vec{\varepsilon}\,'^* \cdot \hat{q} \,
\vec{\varepsilon} \cdot \hat{q}' \,
\vec{\sigma} \cdot \hat{e}_{y} \, b_{2}^{t\prime}
\nonumber\\
& & \quad \; + i \, \vec{\varepsilon}\,'^* \cdot \hat{q} \,
\vec{\varepsilon} \cdot ( \hat{q} \times \hat{q}' ) \,
\vec{\sigma} \cdot \hat{e}_{z} \, b_{3}^{t}
+ i \, \vec{\varepsilon} \cdot \hat{q}' \,
\vec{\varepsilon}\,'^* \cdot ( \hat{q} \times \hat{q}' ) \,
\vec{\sigma} \cdot \hat{e}_{z} \, b_{3}^{t \prime} \big] \chi_{i}\, ,
\end{eqnarray}
where $\chi_{i}$ and $\chi_{f}$ are the Pauli spinors of the nucleon.
There appear 3 spin--independent and 9 spin--dependent amplitudes
as functions of 3 variables, e.g., the covariants 
$q^{2}, \; q \cdot q'$ and $q \cdot p$.
In the following we will use an equivalent, non--covariant 
set of arguments, $\omega'$, $\bar{q}$ and 
$\cos \vartheta = \hat{q} \cdot \hat{q}'$.
The orthonormal coordinate system
 \begin{equation} \label{gl_7}
 \hat{e}_{x} = \frac{\hat{q}' - \cos\vartheta \hat{q}} {\sin\vartheta}\,, \;
 \hat{e}_{y} = \frac{\hat{q} \times \hat{q}'} {\sin\vartheta}\,, \;
 \hat{e}_{z} = \hat{q} 
 \end{equation}
is fixed by the momenta of the photons.

\section{Generalized Polarizabilities}
\label{kap_3}

The scattering amplitude is decomposed into the Born contribution 
$T_{B}$ and the non--Born or residual contribution $T_{R}$,
\begin{equation} \label{gl_8}
T = T_{B} + T_{R} \,,
\end{equation}
and the generalized polarizabilities are given by a low energy expansion
of $T_{R}$ with respect to $\omega'$ \cite{Guichon_95}.
As has been pointed out in Refs. \cite{Guichon_95,Scherer_96}, the
splitting in (\ref{gl_8}) is not unique.
Contributions which are regular in the limit $\omega' \to 0$ and 
separately gauge invariant can be shifted from the Born amplitude to the
residual amplitude and vice versa.
Therefore, when calculating the generalized polarizabilities, one has
to specify which Born terms have been subtracted from the full amplitude,
since different Born terms lead to different numerical values of the
polarizabilities. 
In our calculation we use the Born amplitude as defined in \cite{Metz_96a}.

To be specific, the generalized polarizabilities have been derived from
the multipoles $H_{R}^{(\rho' L' , \rho L)S}(\omega' , \bar{q})$ of the
residual amplitude \cite{Guichon_95}.
In this notation $\rho \, (\rho')$ represents the charge $(0)$,
magnetic $(1)$ or electric $(2)$ nature of the initial (final) photon,
while $L \, (L')$ refers to its angular momentum.
The quantum number $S$ characterizes the no spin--flip $(S = 0)$ and the
spin--flip $(S = 1)$ transitions.

For the definition of the generalized polarizabilities it is necessary
to know the low energy behaviour of the multipoles if 
$(\omega' , \bar{q}) \to (0,0)$.
While the Coulomb and the magnetic transitions are well behaved in that
limit, the electric transitions of the virtual photon depend on the path
along which the origin in the $\omega'$-$\bar{q}$-plane is approached.
Since we are only interested in the leading order term in 
$| \, \vec{q}\,' | = \omega'$, we can express the electric transitions 
of the outgoing photon by the Coulomb transitions \cite{Guichon_95},
\begin{equation} \label{gl_9}
H_{R}^{(2 L' , \rho L)S} (\omega' , \bar{q}) =
- \sqrt{\frac{L' + 1}{L'}} H_{R}^{(0 L' , \rho L)S} (\omega' , \bar{q})
+ {\mathcal{O}}(\omega'^{L' + 1}) \, .
\end{equation}
The corresponding equation for the incoming photon reads
\begin{equation} \label{gl_10}
H_{R}^{(\rho' L' , 2 L)S} (\omega' , \bar{q}) =
- \sqrt{\frac{L + 1}{L}} \frac{\omega}{\bar{q}}
H_{R}^{(\rho' L' , 0 L)S} (\omega' , \bar{q})
- \sqrt{\frac{2L + 1}{L}}
\hat{H}_{R}^{(\rho' L' , L)S} (\omega' , \bar{q}) \, .
\end{equation}
Note that no expansion in the three--momentum $\bar{q}$ of the virtual
photon is performed.
Hence, one has to keep the difference $\hat{H}_{R}^{(\rho' L' , L)S}$
between the electric and the charge multipoles.
Even if this difference has a mixed multipole character, it has an obvious
and path--independent low energy behaviour \cite{Guichon_95}.

The generalized polarizabilities are defined according to \cite{Guichon_95}
as
\begin{eqnarray} \label{gl_11}
P^{(\rho' L' , \rho L)S} (\bar{q}) & = &
\biggl [ \frac{1}{\omega'^{L'} \bar{q}^{L}}
H_{R}^{(\rho' L' , \rho L)S}(\omega' , \bar{q}) \biggr ] _{\omega' = 0}
\qquad (\rho , \rho' = 0,1) \, ,
\\ \label{gl_12}
\hat{P}^{(\rho' L' , L)S} (\bar{q}) & = &
\biggl [ \frac{1}{\omega'^{L'} \bar{q}^{L+1}}
\hat{H}_{R}^{(\rho' L' , L)S}(\omega' , \bar{q}) \biggr ] _{\omega' = 0}
\qquad (\rho' = 0,1) \, .
\end{eqnarray}
In the following we restrict ourselves to electric and magnetic dipole
transitions of the real photon, which is equivalent to keep only the
leading, linear term in $\omega'$ of the residual amplitude $T_{R}$.
In that case there exist, due to conservation of parity and angular
momentum, 3 scalar multipoles $(S = 0)$ and 7 vector multipoles $(S = 1)$.

To evaluate the vector polarizabilities it is sufficient to treat the
following 5 out of the 9 spin--dependent amplitudes of eqs.
(\ref{gl_5}) and (\ref{gl_6}):
\begin{eqnarray} \label{gl_13}
b_{R,1}^{l} & = & \frac{e^{2}}{4\pi} \sqrt{\frac{E}{m_{N}}} 
\frac{\omega'}{\sin\vartheta} \frac{\sqrt{3}}{2}
\biggl[ \frac{\omega_{0}}{\bar{q}} \cos \vartheta
\Bigl( P^{(11,00)1} 
+ \frac{\bar{q}^{2}}{\sqrt{2}} P^{(11,02)1} \Bigr)
- \sqrt{3} \omega_{0} P^{(01,01)1} \biggr] 
+ {\mathcal{O}}(\omega'^{2}) \, ,
\\ \label{gl_14}
b_{R,3}^{l} & = & \frac{e^{2}}{4\pi} \sqrt{\frac{E}{m_{N}}} 
\omega' \frac{\sqrt{3}}{2}
\biggl[ \frac{\omega_{0}}{\bar{q}} 
\Bigl( - P^{(11,00)1} + \sqrt{2}\bar{q}^{2} P^{(11,02)1} \Bigr) \biggr] 
+ {\mathcal{O}}(\omega'^{2}) \, ,
\\ \label{gl_15}
b_{R,1}^{t} & = & \frac{e^{2}}{4\pi} \sqrt{\frac{E}{m_{N}}}
\frac{\omega'}{\sin\vartheta} \frac{3}{4}
\biggl[ 2 \omega_{0} P^{(01,01)1}
+ \sqrt{2} \bar{q}^{2} \Bigl( P^{(01,12)1} 
+ \sqrt{3} \hat{P}^{(01,1)1} \Bigr) \biggr] 
+ {\mathcal{O}}(\omega'^{2}) \, ,
\\ \label{gl_16}
b_{R,1}^{t\prime} & = & \frac{e^{2}}{4\pi} \sqrt{\frac{E}{m_{N}}} 
\frac{\omega'}{\sin\vartheta} \frac{3}{4}
\biggl[ \bar{q} P^{(11,11)1} 
+ \frac{\sqrt{3}}{\sqrt{2}} \omega_{0} \bar{q} P^{(11,02)1}
+ \frac{\sqrt{5}}{\sqrt{2}} \bar{q}^{3} \hat{P}^{(11,2)1} \biggr] 
+ {\mathcal{O}}(\omega'^{2}) \, ,
\\ \label{gl_17}
b_{R,3}^{t\prime} & = & \frac{e^{2}}{4\pi} \sqrt{\frac{E}{m_{N}}}
\frac{\omega'}{\sin^{2}\vartheta} \frac{3}{4}
\biggl[ 2 \omega_{0} P^{(01,01)1}
+ \sqrt{2} \bar{q}^{2} \Bigl( - P^{(01,12)1} 
+ \sqrt{3} \hat{P}^{(01,1)1} \Bigr)
\nonumber \\ 
& & \hspace{3cm} + \cos\vartheta \Bigl( \bar{q} P^{(11,11)1}
- \frac{\sqrt{3}}{\sqrt{2}} \omega_{0} \bar{q} P^{(11,02)1}
- \frac{\sqrt{5}}{\sqrt{2}} \bar{q}^{3} \hat{P}^{(11,2)1} \Bigr) \biggr]  
+ {\mathcal{O}}(\omega'^{2}) \, .
\end{eqnarray}
Because of their angular dependence, $b_{R,1}^{l}$ and $b_{R,3}^{t \prime}$
contain two independent linear combinations of the polarizabilities.

Based upon gauge invariance, Lorentz invariance, parity conservation,
and charge conjugation in connection with nucleon crossing, it was shown
in Ref. \cite{Drechsel_97} that only four independent vector polarizabilities
exist.
This fact is reflected by the equations
\begin{eqnarray} \label{gl_18}
0 & = & P^{(11,11)1}(\bar{q})
 + \sqrt{\frac{3}{2}} \omega_0 P^{(11,02)1}(\bar{q})
 + \sqrt{\frac{5}{2}} \bar{q}^2 \hat{P}^{(11,2)1}(\bar{q}) \, ,
\\ \label{gl_19}
0 & = & 2 \omega_0 P^{(01,01)1}(\bar{q})
 + 2 \frac{\bar{q}^2}{\omega_0} P^{(11,11)1}(\bar{q})
 - \sqrt{2} \bar{q}^2 P^{(01,12)1}(\bar{q})
 + \sqrt{6} \bar{q}^2 \hat{P}^{(01,1)1}(\bar{q}) \, ,
\\ \label{gl_20}
0 & = & 3 \frac{\bar{q}^2}{\omega_0} P^{(01,01)1}(\bar{q})
 - \sqrt{3} P^{(11,00)1}(\bar{q}) 
 - \sqrt{\frac{3}{2}} \bar{q}^2 P^{(11,02)1}(\bar{q}) \, .
\end{eqnarray}
Due to (\ref{gl_18}) the amplitude $b_{R,1}^{t \prime}$ vanishes
to linear order in $\omega'$.
Equations (\ref{gl_19},\ref{gl_20}) relate the angular--independent
and the angular--dependent terms of $b_{R,1}^{l}$ and $b_{R,3}^{t \prime}$.
Moreover, the relations
\begin{equation} \label{gl_21}
P^{(01,01)1}(0) = P^{(11,11)1}(0) = P^{(11,00)1}(0) = 0
\end{equation}
have been established \cite{Drechsel_97}.
At $\bar{q} = 0$, further specific relations between the polarizabilities 
and their derivatives can be obtained by expanding 
(\ref{gl_18})--(\ref{gl_20}) with respect to $Q_{0}^{2}$ or $\bar{q}^{2}$.
We have used such expansions up to $Q_{0}^{4}$ in order to test our 
analytical results.
Furthermore we note that the polarizabilities are actually functions 
of $\bar{q}^{2}$ \cite{Drechsel_97}. 

Equivalent to the parametrization of the spin--flip amplitude of VCS
in terms of generalized vector polarizabilities, Ragusa \cite{Ragusa_93}
expressed the spin--flip amplitude in RCS, to lowest order in the photon
energy, by 4 polarizabilities, called 
$\gamma_{1}, \gamma_{2}, \gamma_{3}, \gamma_{4}$.
Two linear combinations of these $\gamma_{i}$ can be related to the 
generalized polarizabilities \cite{Drechsel_97},
\begin{equation} \label{gl_22}
\gamma_{3} = - \frac{e^{2}}{4\pi}\frac{3}{\sqrt{2}} P^{(01,12)1}(0) \,,
\quad
\gamma_{2} + \gamma_{4} = - \frac{e^{2}}{4\pi} \frac{3\sqrt{3}}{2\sqrt{2}}
P^{(11,02)1}(0) \,.
\end{equation}
To recover the remaining two combinations of Ragusa, one has to go beyond 
the leading order term in $\omega'$.
We note that the right hand sides of (\ref{gl_22}) are not unique,
because of the relations between the generalized polarizabilities.

\section{Results and discussion}
\label{kap_4}

\begin{figure}[h] 
\centerline{
\epsfxsize=11.5cm
\epsfbox{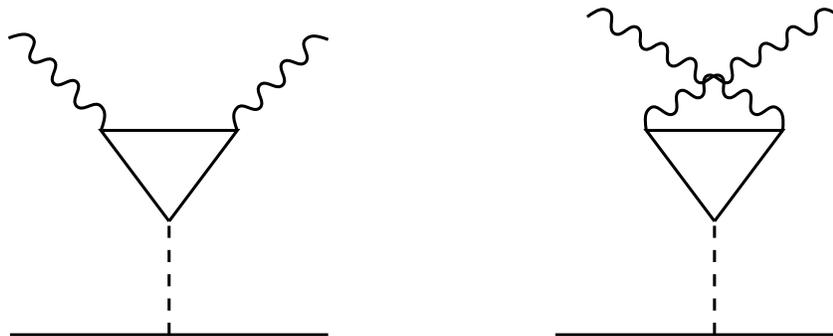}}
\caption{\label{fig_1} 
 Anomaly diagrams for virtual Compton scattering off the nucleon. 
 Solid lines: nucleons, wavy lines: photons, dotted lines: neutral pions.}
\end{figure}

The one--loop diagrams contributing to the generalized polarizabilities 
in the LSM for infinite sigma mass were shown in \cite{Metz_96a}.
While the sigma exchange in the $t$ channel strongly influences the 
numerical values of the scalar polarizabilities $\alpha$
and $\beta$, it does not contribute to the vector polarizabilities.
On the other hand, the $\pi^{0}$ exchange in the $t$ channel (anomaly, 
see Fig. \ref{fig_1}) is irrelevant in the spin--independent 
case, but very important for calculations of the vector polarizabilities.
Since the anomaly is gauge invariant and regular in the soft photon limit,
it could also be considered as part of the Born amplitude \cite{Guichon_95}.
Our results without the anomaly are discussed in the following 
subsection A, and the anomaly is treated in subsection B.

\subsection{Numerical results}
\label{sub_41}
\begin{figure}[h]
\centerline{
\epsfxsize=15.5cm
\epsfbox{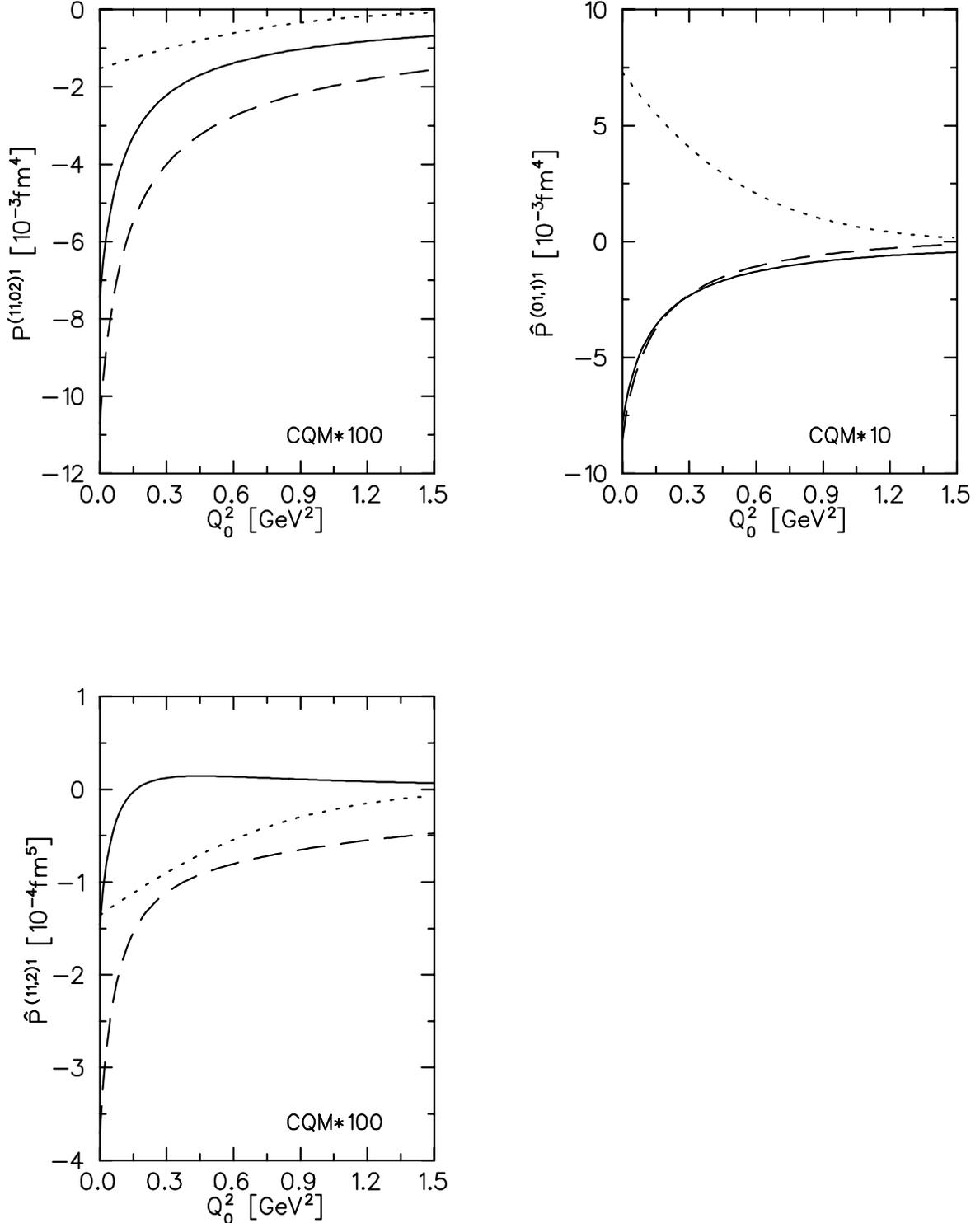}}
\caption{\label{fig_2}
 The vector polarizabilities as functions of momentum transfer.
 Solid line: calculation with the LSM for the proton, dashed line: 
 LSM result for the neutron, dash--dotted line: CQM for the proton
 {\protect\cite{Liu_96}}.
 Note that the CQM results have been scaled.} 
\end{figure}

Figure \ref{fig_2} displays our numerical results for the 7 vector
polarizabilities of both nucleons.
As has been shown analytically, the polarizabilities $P^{(01,01)1}$, 
$P^{(11,11)1}$ and $P^{(11,00)1}$ satisfy the constraints of (\ref{gl_21}).
Moreover, our results fulfill the relations (\ref{gl_18})--(\ref{gl_20}) 
for arbitrary values of $\bar{q}$.

The vector polarizabilities of proton and neutron differ significantly 
in four cases.
For $P^{(01,01)1}$ and $P^{(11,00)1}$ this difference is growing as function 
of $Q_{0}^{2}$, $P^{(11,02)1}$ and $\hat{P}^{(11,2)1}$ differ for the two
nucleons over the whole range of $Q_{0}^{2}$.

In the following discussion we focus on the proton.
Apart from a sign the values of $P^{(01,01)1}$ and $P^{(11,11)1}$ are 
very similar, even if their behaviour on $Q_{0}^{2}$ is slightly different.
It is worthwhile to compare these two polarizabilities to their scalar
partners $P^{(01,01)0}$ and $P^{(11,11)0}$ (see the results in ref. 
\cite{Metz_96a}), which are proportional to $\alpha$ and $\beta$, 
respectively.
While the absolute values of the scalar polarizabilities decrease as
function of $Q_{0}^{2}$, the vector polarizabilities increase.
On the other hand, the ratios evaluated at 
$Q_{0}^{2} = 0.33 \, \text{GeV}^{2}$,
$P_{p}^{(01,01)0} / P_{p}^{(01,01)1} \approx - 9.4 \,,$
$P_{p}^{(11,11)0} / P_{p}^{(11,11)1} \approx - 7.8 \,,$
demonstrate that the scalar polarizabilities $\alpha$ and $\beta$ will
dominate the residual amplitude for the kinematic typical at MAMI.
Contrary to this, the remaining scalar polarizability
$\hat{P}_{p}^{(01,1)0}$ is comparable to the vector polarizability
$\hat{P}_{p}^{(01,1)1}$.
For increasing $Q_{0}^{2}$ both quantities tend to zero but remain of the 
same order of magnitude.

Among the 3 polarizabilities $P^{(01,12)1}$, $P^{(11,02)1}$ and
$\hat{P}^{(01,1)1}$ with the common unit $\text{fm}^{4}$, 
$P^{(01,12)1}$ is generally suppressed.

Since the polarizabilities have different dimensions due to their definitions
(\ref{gl_11},\ref{gl_12}), the amplitudes of (\ref{gl_13})--(\ref{gl_17})
are constructed by multiplying the polarizabilities with different 
kinematical factors.
Therefore, the influence of a particular polarizability can only be seen 
after evaluating the expansion coefficients (\ref{gl_13})--(\ref{gl_17}).
 
The LSM and the CQM calculation of Liu et al. \cite{Liu_96} predict 
signs different from ours for two of the vector polarizabilities
($P^{(01,01)1}$, $\hat{P}^{(01,1)1}$).
Furthermore, with the exception of $P^{(11,00)1}$, the variation of the 
vector polarizabilities at low $Q_{0}^{2}$ is much stronger in the LSM 
than in the CQM, as was the case for the scalar polarizabilities
\cite{Metz_96a}. 
However, the most remarkable difference between the models is in the 
absolute values of the vector polarizabilities.
Except for $P^{(11,11)1}$ and $P^{(01,12)1}$, our results are substantially
larger than the CQM predictions, in the most striking case of
$P^{(11,00)1}$ by three orders of magnitude.
This result indicates that for most of the vector polarizabilities the
nonresonant background, described in our calculation to one--loop, is more
important than the nucleon resonances.

The most obvious resonance contribution is due to the strong M1 transition
to the $\Delta(1232)$, which leads to the large value of $P^{(11,11)1}$
in the CQM.
On the other hand, the CQM violates the model--independent constraint
$P^{(11,11)1}(0) = 0$.
This shortcoming of the CQM can probably be traced back to the lack of
covariance.
Similarly, the main contribution to $P^{(01,12)1}$ in the CQM is caused by
the $D_{13}(1520)$, which is clearly visible in the photoabsorption
spectrum.

\subsection{Contribution of anomaly diagrams}
\label{sub_42}
\begin{figure}[h]
\centerline{
\epsfxsize=13cm
\epsfbox{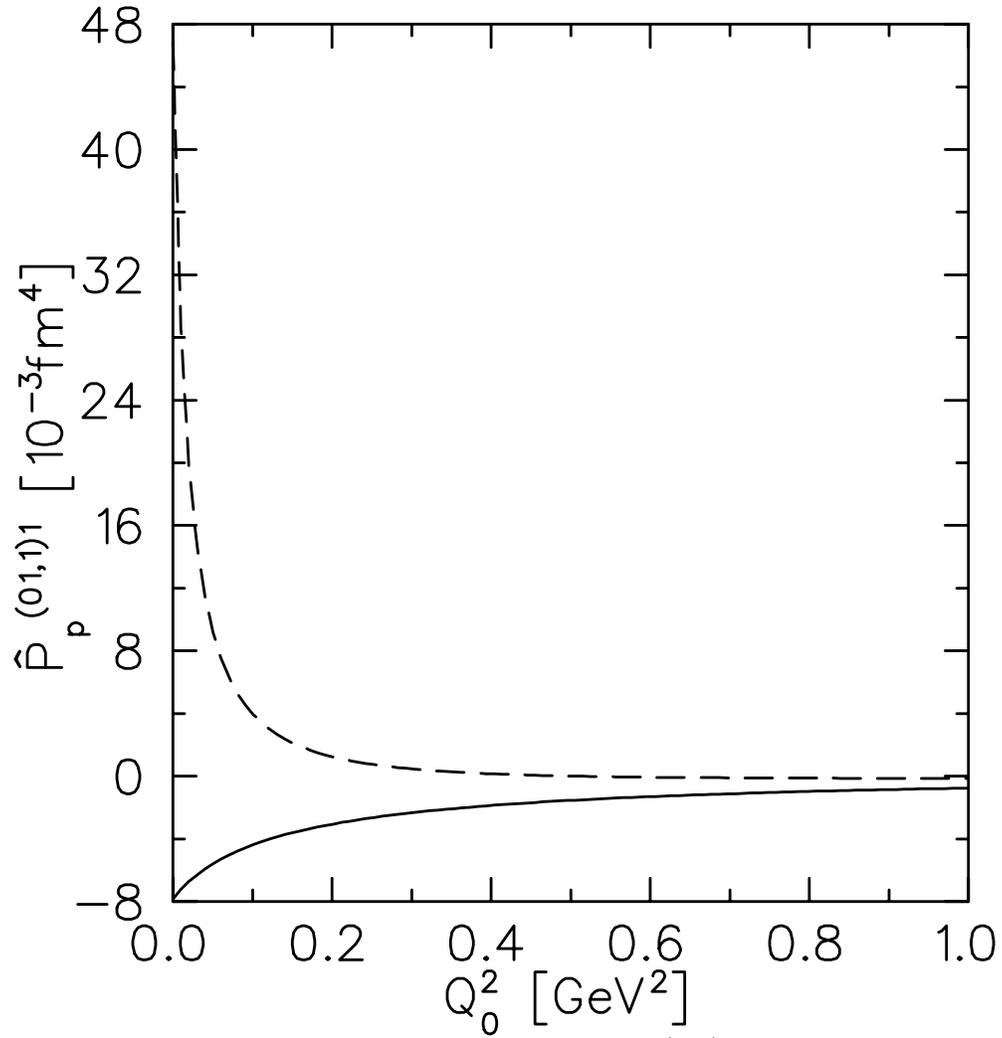}}
\caption{\label{fig_3}
 Influence of the anomaly on the polarizability $\hat{P}_{p}^{(01,1)1}$.
 Solid line: without anomaly, dashed line: anomaly included.} 
\end{figure}

The two anomaly diagrams in Fig. \ref{fig_1} give rise to the amplitude  
\begin{eqnarray} \label{gl_23}
T_{a} & = & \frac{e^{2} g_{\pi N}^{2}}{2\pi} 
 \frac{m_{N}}{t - m_{\pi}^{2}}
 \tau_{0} \bar{u}(p') \gamma_{5} u(p) \textrm{Tr}
 (\gamma_{5} \esla \qsla \esla' \qsla') L_{0}(t,q^{2}) \,,
\\ 
L_{0}(t,q^{2}) & = & \frac{1}{16\pi^{2}} \int_{0}^{1} dx \int_{0}^{1} dy 
 \frac{x}{m_{N}^{2} - tx^{2}y(1-y) - q^{2}x(1-x)y} \,,
\nonumber
\end{eqnarray}
with the Mandelstam variable $t = (q-q')^{2}$.
The matrix $\tau_{0}$ acting in isospace gives opposite signs
for proton and neutron.
The anomaly contributes to 5 of the 7 vector polarizabilities.
In the case of the proton these contributions read
\begin{eqnarray} \label{gl_24}
& & P_{p,a}^{(11,00)1}(Q_{0}^{2}) = 
\frac{1}{\sqrt{3}} \frac{\omega_{0}}{m_{N}} I(Q_{0}^{2})\,, 
\qquad \quad \;\;
P_{p,a}^{(11,11)1}(Q_{0}^{2}) = 
\frac{1}{m_{N}} \frac{\omega_{0}^{2}}{\bar{q}^{2}} I(Q_{0}^{2}) \,, 
\nonumber \\
& & P_{p,a}^{(01,12)1}(Q_{0}^{2}) = 
\frac{1}{\sqrt{2}m_{N}^{2}} \frac{m_{N}\omega_{0}}{\bar{q}^{2}} 
 I(Q_{0}^{2})\,, \quad
P_{p,a}^{(11,02)1}(Q_{0}^{2}) = 
- \frac{\sqrt{2}}{\sqrt{3}m_{N}^{2}} \frac{m_{N}\omega_{0}}{\bar{q}^{2}} 
 I(Q_{0}^{2}) \,, 
\nonumber \\
& & \hat{P}_{p,a}^{(01,1)1}(Q_{0}^{2}) =
- \frac{1}{\sqrt{6}m_{N}^{2}} \frac{m_{N}\omega_{0}}{\bar{q}^{2}}
 I(Q_{0}^{2}) \,,
\\ \nonumber
& & \textrm{with} \; I(Q_{0}^{2}) = \frac{g_{\pi N}^{2}}{6\pi^{2}m_{N}^{2}}
\sqrt{\frac{E + m_{N}}{2E}} \frac{m_{N}^{2}}{m_{\pi}^{2} + Q_{0}^{2}}
\frac{m_{N}^{2}}{\sqrt{4m_{N}^{2}Q_{0}^{2} + Q_{0}^{4}}}
\ln \frac{2m_{N}^{2} + Q_{0}^{2} + \sqrt{4m_{N}^{2}Q_{0}^{2} + Q_{0}^{4}}}
 {2m_{N}^{2} + Q_{0}^{2} - \sqrt{4m_{N}^{2}Q_{0}^{2} + Q_{0}^{4}}} \,,
\\ \nonumber
& & \textrm{and} \; I(0) = \frac{g_{\pi N}^{2}}{6\pi^{2}m_{\pi}^{2}} \;.
\nonumber
\end{eqnarray}
In the 3 linear combinations of polarizabilities (structure functions) 
that can be determined in an unpolarized experiment, the contribution 
of the anomaly drops out.
However, in the case of double polarization experiments the anomaly 
is very essential.
To demonstrate the great importance of the anomaly we have, as an
example, plotted the polarizability $\hat{P}_{p}^{(01,1)1}$ with and without 
anomaly contribution in Fig. \ref{fig_3}.

\subsection{Analytical results}
As in the scalar case \cite{Metz_96a} all vector polarizabilities can be
calculated analytically at $Q_{0}^{2} = 0$.
Instead of quoting the complete analytical expressions we have given an
expansion in powers of the pion mass.
Using the definition $\mu = m_{\pi}/m_{N}$ we arrive at the
following results:
\begin{eqnarray}  \label{gl_25}
P_{p}^{(11,02)1}(0) & = & 
 \sqrt{\frac{2}{3}} C
 \Biggl[ \Bigl( \frac{6}{\mu^{2}} \Bigr) - \frac{1}{2\mu^{2}} 
        + \frac{3 \pi}{8 \mu} + 3\ln\mu + \frac{13}{4} 
 + {\mathcal{O}}(\mu) \Biggr] \, ,
 \nonumber \\
 & & \vphantom{.} \nonumber \\
P_{n}^{(11,02)1}(0) & = & 
 \sqrt{\frac{2}{3}} C
 \Biggl[ \Bigl( - \frac{6}{\mu^{2}} \Bigr) - \frac{1}{2\mu^{2}} 
        - \frac{\pi}{8 \mu} + \ln\mu + \frac{3}{4} 
 + {\mathcal{O}}(\mu) \Biggr] \, ,
 \nonumber\\
 & & \vphantom{.} \nonumber \\
P_{p}^{(01,12)1}(0) & = & 
 \frac{1}{\sqrt{2}} C
 \Biggl[ \Bigl( - \frac{6}{\mu^{2}} \Bigr) - \frac{1}{2\mu^{2}} 
        + \frac{7 \pi}{8 \mu} + 3\ln\mu - \frac{5}{4} 
 + {\mathcal{O}}(\mu) \Biggr] \, ,
 \nonumber\\
 & & \vphantom{.} \nonumber \\
P_{n}^{(01,12)1}(0) & = & 
 \frac{1}{\sqrt{2}} C
 \Biggl[ \Bigl( \frac{6}{\mu^{2}} \Bigr) - \frac{1}{2\mu^{2}} 
        + \frac{5 \pi}{8 \mu} + \ln\mu - \frac{1}{4} 
 + {\mathcal{O}}(\mu) \Biggr] \, ,
 \nonumber\\
 & & \vphantom{.} \nonumber \\
\hat{P}_{p}^{(01,1)1}(0) & = & 
 \frac{1}{\sqrt{6}} C
 \Biggl[ \Bigl( \frac{6}{\mu^{2}} \Bigr) - \frac{3}{2\mu^{2}} 
        + \frac{21 \pi}{8 \mu} + 21\ln\mu + \frac{85}{4} 
 + {\mathcal{O}}(\mu) \Biggr] \, ,
 \nonumber\\
 & & \vphantom{.} \nonumber \\
\hat{P}_{n}^{(01,1)1}(0) & = & 
 \frac{1}{\sqrt{6}} C
 \Biggl[ \Bigl( - \frac{6}{\mu^{2}} \Bigr) - \frac{3}{2\mu^{2}} 
        + \frac{15 \pi}{8 \mu} + 7\ln\mu + \frac{1}{4} 
 + {\mathcal{O}}(\mu) \Biggr] \, ,
 \nonumber\\
 & & \vphantom{.} \nonumber \\
\hat{P}_{p}^{(11,2)1}(0) & = & 
 \sqrt{\frac{2}{5}} \frac{C}{m_{N}}
 \Biggl[ - \frac{\pi}{4 \mu} - 3\ln\mu - 4 
 + {\mathcal{O}}(\mu) \Biggr] \, ,
 \nonumber\\
 & & \vphantom{.} \nonumber \\
\hat{P}_{n}^{(11,2)1}(0) & = & 
 \sqrt{\frac{2}{5}} \frac{C}{m_{N}}
 \Biggl[ - \frac{3\pi}{8 \mu} - \ln\mu + \frac{1}{4} 
 + {\mathcal{O}}(\mu) \Biggr] \, ,
 \\
 & & \vphantom{a} \nonumber \\
 & & \textrm{with} \quad C = \frac{g_{\pi N}^{2}}{72\pi^{2}m_{N}^{4}} \,.
 \nonumber
\end{eqnarray} 
In eq. (\ref{gl_25}) the expressions in round brackets denote the 
contribution of the anomaly.
The results confirm that the anomaly diagrams dominate the 
polarizabilities whenever they contribute.
With the exception of $\hat{P}^{(11,2)1}$, the scalar and vector 
polarizabilities show two common properties:
First, the leading terms of the loop contributions do not depend on 
isospin, and second, the dominant chiral logarithm is three times larger for 
the proton than for the neutron.
However, the scalar and vector polarizabilities differ in their chiral 
behaviour.
Apart from $\hat{P}^{(11,2)1}$, the leading term of the non--vanishing
vector polarizabilities is proportional to $m_{\pi}^{-2}$, while the 
scalar polarizabilities diverge like $m_{\pi}^{-1}$ in the chiral limit
\cite{Metz_96a}.

By means of eq. (\ref{gl_22}) one immediately obtains the third order
spin--polarizabilities $\gamma_{3}$ and $\gamma_{2} + \gamma_{4}$.
The leading order term of our results agrees with a calculation
of Bernard et al. \cite{Bernard_95} who evaluated all 4 
spin--polarizabilities for RCS on the basis of heavy baryon chiral 
perturbation theory (HBChPT) to lowest order (${\cal{O}}(p^{3}))$.

The relationship between the LSM and ChPT holds for all vector 
polarizabilities.
A calculation of Hemmert et al. in HBChPT to the order ${\cal{O}}(p^{3})$ 
\cite{Hemmert_97a} completely agrees with the leading terms of the 
chiral expansion (\ref{gl_25}), except for $\hat{P}^{(11,2)1}$,
which vanishes in ChPT to order ${\cal{O}}(p^{3})$.
Therefore, in ChPT a ${\cal{O}}(p^{4})$ calculation is required    
to obtain the terms in $m_{\pi}^{-1}$.

The agreement between HBChPT and the LSM is restricted to the leading 
$m_{\pi}^{-2}$ term of the chiral expansion.
The next to leading order contributions will be modified by various 
low energy constants which enter a ${\cal{O}}(p^{4})$ calculation
and describe physics beyond the scope of the sigma model (e.g. resonance
contributions, kaon--loops).
As an example we refer to the calculation of $\alpha$ and $\beta$ in RCS to
the order ${\cal{O}}(p^{4})$ in HBChPT \cite{Bernard_93}.

At $Q_{0}^{2} = 0$, all vector polarizabilities have a finite
derivative whose leading terms in the chiral expansion are given in
Appendix A.
The derivatives of those polarizabilities which vanish at $Q_{0}^{2} = 0$ 
diverge like $m_{\pi}^{-2}$ in the chiral limit, the slope of the 
remaining polarizabilities is proportional to $m_{\pi}^{-4}$.
All analytical results given in App. A satisfy the model--independent 
relations (\ref{gl_18})--(\ref{gl_20}) between the vector polarizabilities.
We stress that we have obtained a complete agreement with the HBChPT 
calculation to order ${\cal{O}}(p^{3})$ \cite{Hemmert_97a}, also 
for the first and second derivatives of the polarizabilities.

\section{Summary and Conclusion}
\label{kap_5}
The non Born amplitude of VCS off the nucleon may be parametrized
by 10 generalized polarizabilities, to leading order in the 
{\it cm} energy \cite{Guichon_95}.
To complete our previous calculation \cite{Metz_96a} we have evaluated 
the 7 vector polarizabilities which determine the spin--dependent 
amplitude of the VCS reaction.
To this end, we used the LSM in the one--loop approximation and in the 
limit of an infinite sigma mass.
Both our numerical and analytical results are in complete agreement with
3 relations between the vector polarizabilities which have been derived 
in a model--independent way \cite{Drechsel_97}.

We find that the anomaly diagrams strongly affect 5 of the vector 
polarizabilities. 
In particular, when measuring double polarization observables like in
the proposed reaction $p(\vec{e},e'\vec{p}\,)\gamma$ 
\cite{Vanderhaeghen_97b}, the anomaly contributions become quite important.
In the 3 independent structure functions of the unpolarized cross section, 
the anomaly contributions cancel exactly.

At $Q_{0}^{2} = 0$, we have performed a Taylor expansion of all 
polarizabilities (with respect to $Q_{0}^{2}$) keeping the first two 
non--vanishing coefficients.
The various Taylor coefficients have been expanded in powers of $m_{\pi}$
and compared with the results of ChPT.
It turns out that both calculations totally agree in the leading, 
isospin--independent term of the chiral expansion \cite{Hemmert_97a}.
The same is true in the case of the third order spin--polarizabilities 
$\gamma_{3}$ and $\gamma_{2} + \gamma_{4}$ of real Compton scattering
\cite{Bernard_95}.
Of course, a future ${\cal{O}}(p^{4})$ calculation in ChPT will 
give rise to additional low energy constants.

In comparison to the CQM \cite{Liu_96} the LSM leads to substantially 
larger results for 5 of the 7 vector polarizabilities. 
In the case of $P^{(11,00)1}$ the ratio of the predictions reaches three
orders of magnitude.
We conclude that most of the vector polarizabilities are dominated by 
virtual excitations of nonresonant $\pi N$ scattering states. 
Only in the case of $P^{(11,11)1}$, the result of the LSM is small 
compared to the CQM, which is due to the strong influence of the
$\Delta(1232)$ resonance on that polarizability.
The CQM, however, does not fulfill the model--independent condition
$P^{(11,11)1}(0) = 0$. 

In conclusion, the generalized vector polarizabilities represent suitable
observables to distinguish between different models, and, in
particular, to test the chiral structure of the nucleon.
Accordingly, experimental effort to measure these quantities is certainly
worthwhile, despite the huge difficulties involved in clearly separating
the effects of individual polarizabilities.

\section{Acknowledgement}
We would like to thank G. Kn\"ochlein and S. Scherer for several useful
discussions.
We are also grateful to G. Q. Liu for providing us with the results of the
quark model calculation.
This work has been supported by the Deutsche Forschungsgemeinschaft 
(SFB 201).

\appendix
\section{Chiral expansion of the first and second derivatives of the
 polarizabilities}
Below we list the chiral expansion of the first derivatives of the vector
polarizabilities.
The constant $C$ is defined in equation (\ref{gl_25}).
\begin{eqnarray} \label{gl_26}
\frac{d}{d Q_{0}^{2}} P_{p}^{(01,01)1}(0) & = & 
 \frac{C}{2m_{\pi}}
 \Biggl[ \frac{1}{\mu} - \frac{17\pi}{8}
 + {\mathcal{O}}(\mu) \Biggr] \,,
 \nonumber \\
 & & \vphantom{.} \nonumber \\
\frac{d}{d Q_{0}^{2}} P_{n}^{(01,01)1}(0) & = & 
 \frac{C}{2m_{\pi}}
 \Biggl[ \frac{1}{\mu} - \frac{9\pi}{8}
  + {\mathcal{O}}(\mu) \Biggr] \,,
 \nonumber\\
 & & \vphantom{.} \nonumber \\
\frac{d}{d Q_{0}^{2}} P_{p}^{(11,11)1}(0) & = & 
 \frac{C}{2m_{\pi}}
 \Biggl[ \Bigl( \frac{6}{\mu} \Bigr) - \frac{1}{2\mu} + \frac{7\pi}{8}
 + {\mathcal{O}}(\mu) \Biggr] \,,
 \nonumber\\
 & & \vphantom{.} \nonumber \\
\frac{d}{d Q_{0}^{2}} P_{n}^{(11,11)1}(0) & = &
 \frac{C}{2m_{\pi}}
 \Biggl[ \Bigl( - \frac{6}{\mu} \Bigr) - \frac{1}{2\mu} + \frac{5\pi}{8}
 + {\mathcal{O}}(\mu) \Biggr] \,,
 \nonumber\\
 & & \vphantom{.} \nonumber \\
\frac{d}{d Q_{0}^{2}} P_{p}^{(11,00)1}(0) & = & 
 \frac{1}{\sqrt{3}} \frac{Cm_{N}}{2m_{\pi}}
 \Biggl[ \Bigl( - \frac{12}{\mu} \Bigr) - \frac{5}{\mu} + 12\pi
 + {\mathcal{O}}(\mu) \Biggr] \,,
 \nonumber\\
 & & \vphantom{.} \nonumber \\
\frac{d}{d Q_{0}^{2}} P_{n}^{(11,00)1}(0) & = &
 \frac{1}{\sqrt{3}} \frac{Cm_{N}}{2m_{\pi}}
 \Biggl[ \Bigl( \frac{12}{\mu} \Bigr) - \frac{5}{\mu} + 7\pi
 + {\mathcal{O}}(\mu) \Biggr] \,,
 \nonumber\\
 & & \vphantom{.} \nonumber \\
\frac{d}{d Q_{0}^{2}} P_{p}^{(11,02)1}(0) & = &
 \sqrt{\frac{2}{3}} \frac{2Cm_{N}}{5m_{\pi}^{3}}
 \Biggl[ \Bigl( - \frac{15}{\mu} \Bigr) + \frac{1}{4\mu} 
        - \frac{3\pi}{16}
 + {\mathcal{O}}(\mu) \Biggr] \,,
 \nonumber\\
 & & \vphantom{.} \nonumber \\
\frac{d}{d Q_{0}^{2}} P_{n}^{(11,02)1}(0) & = & 
 \sqrt{\frac{2}{3}} \frac{2Cm_{N}}{5m_{\pi}^{3}}
 \Biggl[ \Bigl( \frac{15}{\mu} \Bigr) + \frac{1}{4\mu} 
        - \frac{\pi}{16}
 + {\mathcal{O}}(\mu) \Biggr] \,,
 \nonumber\\
 & & \vphantom{.} \nonumber \\
\frac{d}{d Q_{0}^{2}} P_{p}^{(01,12)1}(0) & = &
 \frac{1}{\sqrt{2}} \frac{2Cm_{N}}{5m_{\pi}^{3}}
 \Biggl[ \Bigl( \frac{15}{\mu} \Bigr) + \frac{1}{4\mu} 
        - \frac{9\pi}{32}
 + {\mathcal{O}}(\mu) \Biggr] \,,
 \nonumber\\
 & & \vphantom{.} \nonumber \\
\frac{d}{d Q_{0}^{2}} P_{n}^{(01,12)1}(0) & = &
 \frac{1}{\sqrt{2}} \frac{2Cm_{N}}{5m_{\pi}^{3}}
 \Biggl[ \Bigl( - \frac{15}{\mu} \Bigr) + \frac{1}{4\mu} 
        - \frac{7\pi}{32}
 + {\mathcal{O}}(\mu) \Biggr] \,,
 \nonumber\\
 & & \vphantom{.} \nonumber \\
\frac{d}{d Q_{0}^{2}} \hat{P}_{p}^{(01,1)1}(0) & = &
 \frac{1}{\sqrt{6}} \frac{2Cm_{N}}{5m_{\pi}^{3}}
 \Biggl[ \Bigl( - \frac{15}{\mu} \Bigr) + \frac{1}{2\mu} 
        - \frac{9\pi}{16}
 + {\mathcal{O}}(\mu) \Biggr] \,,
 \nonumber\\
 & & \vphantom{.} \nonumber \\
\frac{d}{d Q_{0}^{2}} \hat{P}_{n}^{(01,1)1}(0) & = & 
 \frac{1}{\sqrt{6}} \frac{2Cm_{N}}{5m_{\pi}^{3}}
 \Biggl[ \Bigl( \frac{15}{\mu} \Bigr) + \frac{1}{2\mu} 
        - \frac{7\pi}{16}
 + {\mathcal{O}}(\mu) \Biggr] \,,
 \nonumber\\
 & & \vphantom{.} \nonumber \\
\frac{d}{d Q_{0}^{2}} \hat{P}_{p}^{(11,2)1}(0) & = &
 \sqrt{\frac{2}{5}} \frac{3C}{10m_{\pi}^{3}}
 \Biggl[ \frac{1}{12 \mu} - \frac{\pi}{32}
 + {\mathcal{O}}(\mu) \Biggr] \,,
 \nonumber\\
 & & \vphantom{.} \nonumber \\
\frac{d}{d Q_{0}^{2}} \hat{P}_{n}^{(11,2)1}(0) & = &
 \sqrt{\frac{2}{5}} \frac{3C}{10m_{\pi}^{3}}
 \Biggl[ \frac{1}{12 \mu} + \frac{\pi}{32}
 + {\mathcal{O}}(\mu) \Biggr] \,.
\end{eqnarray}
We also quote the second derivatives of those polarizabilities which
vanish at $Q_{0}^{2}=0$.
\begin{eqnarray} \label{gl_27}
\frac{d^{2}}{(d Q_{0}^{2})^{2}} P_{p}^{(01,01)1}(0) & = &
 \frac{2C}{5m_{\pi}^{3}}
 \Biggl[ - \frac{3}{8 \mu} + \frac{15\pi}{32}
 + {\mathcal{O}}(\mu) \Biggr] \,,
 \nonumber \\
 & & \vphantom{.} \nonumber \\
\frac{d^{2}}{(d Q_{0}^{2})^{2}} P_{n}^{(01,01)1}(0) & = &
 \frac{2C}{5m_{\pi}^{3}}
 \Biggl[ - \frac{3}{8 \mu} + \frac{9\pi}{32}
 + {\mathcal{O}}(\mu) \Biggr] \,,
 \nonumber\\
 & & \vphantom{.} \nonumber \\
\frac{d^{2}}{(d Q_{0}^{2})^{2}} P_{p}^{(11,11)1}(0) & = &
 \frac{2C}{5m_{\pi}^{3}}
 \Biggl[ \Bigl( - \frac{15}{\mu} \Bigr) + \frac{1}{8 \mu} 
        - \frac{9\pi}{64} 
 + {\mathcal{O}}(\mu) \Biggr] \,,
 \nonumber\\
 & & \vphantom{.} \nonumber \\
\frac{d^{2}}{(d Q_{0}^{2})^{2}} P_{n}^{(11,11)1}(0) & = &
 \frac{2C}{5m_{\pi}^{3}}
 \Biggl[ \Bigl( \frac{15}{\mu} \Bigr) + \frac{1}{8 \mu} 
        - \frac{7\pi}{64} 
 + {\mathcal{O}}(\mu) \Biggr] \,,
 \nonumber\\
 & & \vphantom{.} \nonumber \\
\frac{d^{2}}{(d Q_{0}^{2})^{2}} P_{p}^{(11,00)1}(0) & = &
 \frac{1}{\sqrt{3}} \frac{2Cm_{N}}{5m_{\pi}^{3}}
 \Biggl[ \Bigl( \frac{30}{\mu} \Bigr) + \frac{7}{4 \mu} 
        - \frac{39 \pi}{16} 
 + {\mathcal{O}}(\mu) \Biggr] \,,
 \nonumber\\
 & & \vphantom{.} \nonumber \\
\frac{d^{2}}{(d Q_{0}^{2})^{2}} P_{n}^{(11,00)1}(0) & = &
 \frac{1}{\sqrt{3}} \frac{2Cm_{N}}{5m_{\pi}^{3}}
 \Biggl[ \Bigl( - \frac{30}{\mu} \Bigr) + \frac{7}{4 \mu} 
        - \frac{25 \pi}{16} 
 + {\mathcal{O}}(\mu) \Biggr] \,.
\end{eqnarray}

\newpage
\clearpage

\newpage
\begin{figure}[h]
\centerline{
\epsfxsize=15.5cm
\epsfbox{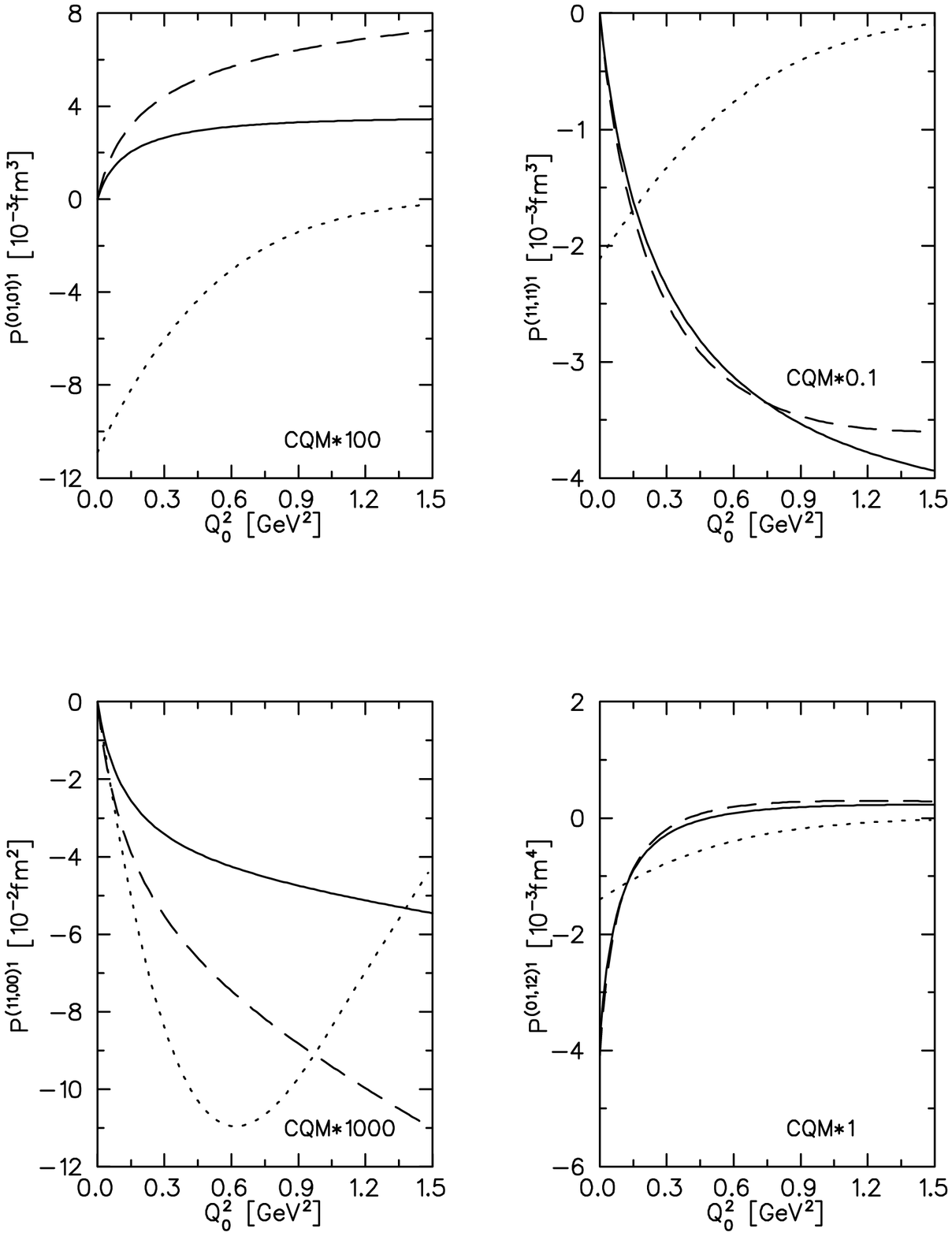}}
\end{figure}

\newpage

\newpage

\end{document}